\documentclass{aa}  
\usepackage[varg]{txfonts}
\usepackage{enumerate, graphicx, sidecap, lipsum}
\usepackage{float}
\usepackage{txfonts}

\usepackage[usenames]{color}
\usepackage[dvipsnames]{xcolor}
\usepackage{hyperref}
\hypersetup{
    colorlinks=true,
    linkcolor=black,
    urlcolor=black,
    citecolor=blue,
    }
\usepackage{marvosym}
\usepackage{csvsimple}
\usepackage{comment}

\usepackage{cancel}
\usepackage[shortlabels]{enumitem}
\usepackage{siunitx}
\DeclareSIUnit \h {\mbox{$h$}}
\DeclareSIUnit \parsec {pc}
\DeclareSIUnit \Msun {\mbox{M$_{\odot}$}}
\DeclareSIUnit \Lsun {\mbox{L$_{\odot}$}}
\DeclareSIUnit \century{century}
\DeclareSIUnit \year{yr}

\newcommand{\sex}{\textsc{SExtractor}}
\newcommand{\psfex}{\textsc{PSFEx}}

\usepackage{natbib}
\bibpunct{(}{)}{;}{a}{}{,} % to follow the A&A style
\begin{document} 

\title{Galaxies in the zone of avoidance: Misclassifications using machine learning tools}
 %\title{The VVV-NIRGC survey IV : miss classifications of galaxies in the Zone of Avoidance using Machine Learning tools.}
%   \title{NIR Analysis of Milky Way X-ray Sources in the Zone of Avoidance using Machine Learning Tools.}
%   \subtitle{Zang et al Revised}

   \author{P. Marchant Cortés\inst{1} \and J. L. Nilo Castell\'on\inst{1} 
  \and M. V. Alonso\inst{2,3} \and L. Baravalle\inst{2,3} \and C. Villalon\inst{2} \and M. A. Sgró\inst{2,3} \and I. V. Daza-Perilla\inst{2,3}   \and M. Soto\inst{4} \and F. Milla Castro\inst{1} \and D. Minniti\inst{5,6,7}  \and N. Masetti\inst{8,5}   \and C. Valotto\inst{2,3}   \and M. Lares\inst{2,3}} 
\institute{  
Departamento Astronom\'ia, Facultad de Ciencias, Universidad de La Serena. Av. Juan Cisternas 1200, La Serena, Chile.
\and
Instituto de Astronom\'{\i}a Te\'orica y Experimental, (IATE-CONICET), Laprida 854, X5000BGR, C\'ordoba, Argentina.
\and
Observatorio Astron\'omico de C\'ordoba, Universidad Nacional de C\'ordoba, Laprida 854, X5000BGR, C\'ordoba, Argentina.
\and 
Instituto de Investigación en Astronomía y Ciencias Planetarias, Universidad de Atacama. Copayapu 485, Copiapó, Chile.
\and
 Instituto de Astrof\'isica, Facultad de Ciencias Exactas, Universidad Andr\'es Bello, Av. Fernandez Concha 700, Las Condes, Santiago, Chile.
 \and
Vatican Observatory, V00120 Vatican City State, Italy.
\and
Departamento de F\'isica, Universidade Federal de Santa Catarina, Trinidade 88040-900, Florianopolis, Brazil.
\and
 INAF - Osservatorio di Astrofisica e Scienza dello Spazio, via Piero Gobetti 101, I-40129 Bologna, Italy.
 }

   \date{Received September 15, 1996; accepted March 16, 1997}

% \abstract{}{}{}{}{} 
% 5 {} token are mandatory
  \abstract
{Automated methods for classifying extragalactic objects in large surveys offer significant advantages compared to manual approaches in terms of efficiency and consistency. However, the existence of the Galactic disk raises additional concerns. These regions are known for high levels of interstellar extinction, star crowding, and limited data sets and studies.} 
  %{} leave it empty if necessary
{In this study, we explore the identification and classification of galaxies in the zone of avoidance (ZoA).  In particular, we compare our results in the near-infrared (NIR) with X-ray data. }
{We analyzed the appearance of objects in the Galactic disk classified as galaxies using a published machine-learning (ML) algorithm and make a comparison with the visually confirmed galaxies from the VVV NIRGC catalog.}
{Our analysis, which includes the visual inspection of all sources cataloged as galaxies throughout the Galactic disk using ML techniques reveals significant differences. Only four galaxies were found in both the NIR and X-ray data sets. Several specific regions of interest within the ZoA exhibit a high probability of being galaxies in X-ray data but closely resemble extended Galactic objects. Our results indicate the difficulty in using ML methods for galaxy classification in the ZoA, which is mainly due to the scarcity of information on galaxies behind the Galactic plane in the training set.  They also highlight the importance of considering specific factors that are present to improve the reliability and accuracy of future studies in this challenging region. }
  {}
\titlerunning{Galaxies in the Zone of Avoidance}
\authorrunning{P. Marchant Cortés}

   \maketitle
%
%________________________________________________________________

\section{Introduction}
Astronomy is moving forward as never before, with much of the progress being driven by the unprecedented amount of data produced by large surveys. New tools have begun to play an essential role in analyzing data, such as machine-learning (ML) algorithms, which have increased the efficiency with which we can identify commonalities across large databases and detect faint and complex patterns.
These algorithms have become a common tool in astronomy because of the large amount of data coming from all-sky surveys, and have been well tested as classifiers for galaxy morphology \citep{Spindler+21}, young stellar object (YSO) finders \citep{Marton+2019}, classifiers of variable stars using light curves \citep{Aguirre+19}, and estimators of photometric redshift \citep{Dainotti+21}, among many others.

All-sky surveys are also increasing in number at different wavelengths.  This improves our understanding of the large-scale evolution and structure of the Universe, the formation of stars and galaxies, and the history of the Milky Way (MW).
At lower Galactic latitudes, the data from these surveys are scarce. This region is known as the zone of avoidance \citep[ZoA;][]{shapley1961review}, 
and became more prominent as complete sky surveys increased in number. It is more critical at $\lvert b\rvert<10^{\circ}$, where extragalactic sources and large-scale structure (LSS) behind the MW are obscured by dust and stellar crowding, dimming the sources by more than $25\%$ in the optical and by about 10\% in the infrared (IR) wavelengths \citep{Henning+1998}.
Classifying extragalactic objects within the ZoA is of critical importance.  We have the opportunity to minimize this gap and  explore the Local Universe in increasing detail. 
This endeavor is pivotal in defining the cosmography of the nearby Universe, which sheds light on the dynamics of the Local Group, giving us insights into the Universe at larger scales. 
It will also allow us to decipher various cosmological parameters, including the peculiar velocity of the Local Group, which exhibits a profound discrepancy relative to the cosmic microwave background  dipole as seen in \cite{Loeb+08(LSScontraints)}.
In order to investigate the LSS, it is crucial to obtain a more complete redshift catalog to fill the ZoA gap in the distribution of the largest mass concentrations in the Local Universe, such as the Great Attractor \citep{Kraan-Korteweg+1996}, the Perseus-Pisces Supercluster \citep{Ramatsoku+2016}, and the Vela Supercluster \citep{Kraan-Korteweg+2017}.

The use of near-infrared (NIR) wavelengths together with radio and X-rays has led to a new wave of extragalactic studies in this region. With improvements in NIR cameras, it has been possible to enlarge the number of discoveries of extragalactic  sources. The first photometric galaxy catalog in the ZoA was provided by the Two Micron All-Sky Survey (2MASS Extended survey, \citealt{2MASS+2006}), which was carried out in order to collect the radial velocities  of these galaxies \citep{Macri+ZoA}.

More recently, the VISTA Variables in the Vía Láctea \citep[VVV;][]{Minniti-VVV} survey also covered these regions. VVV is an ESO public photometric variability survey designed to study the stellar population of the MW bulge and disk in the  $Z$ ($0.87\,\si{\micro\meter}$), $Y$  ($1.02\,\si{\micro\meter}$), $J$ ($1.25\,\si{\micro\meter}$), $H$ ($1.64\,\si{\micro\meter}$), and $K_s$ ($2.14\,\si{\micro\meter}$) NIR passbands. The survey was carried out using the Visible and Infrared Survey Telescope for Astronomy \citep[VISTA;][]{Emerson+2004,Emerson+2006, Emerson+Sutherland2010} 4m telescope at ESO, which is equipped with the VISTA InfraRed CAMera (VIRCAM), a wide-field NIR camera with a pixel scale of 0.34 $\mathrm{arcsec}/\mathrm{pixel}$. The survey covers $300$ $\text{degrees}^{2}$ in the Galactic bulge ($-10^{\circ}\leq\ell\leq 10^{\circ}$ ; $-10^{\circ}\leq b \leq 5^{\circ}$) and $220$ $\text{degrees}^{2}$ of the Galactic disk ($295^{\circ}\leq\ell\leq 350^{\circ}$; $-2^{\circ} \leq b\leq 2^{\circ}$). With a typical limiting $K_s$ magnitude of $17-18$ mag and exceptional data quality, it is the deepest existing data set to explore the LSS in the ZoA. Using the VVV data (\citealt{Baravalle+2021} and references therein), the galaxies behind the Galactic disk were selected by color cuts and visually inspection. The volume of data from the ZoA notably increased with the extension of the VVV survey, known as the VVVX (see Table 1 in \cite{Daza+2023}), and the need to apply ML techniques became evident.
In this sense, \cite{Daza+2023} applied ML algorithms to the northern part of the disk from the VVVX ($10^{\circ}<\ell < 20^{\circ}$, $-4.5^{\circ}<b<+4.5^{\circ}$) survey using the visual classification already performed in the southern part (VVV survey) as a training set. These authors used two samples of data, one based on the NIR images and the other based on the photometric-morphological information obtained mainly from \sex\ \citep{SExtractor}. These samples  were used to separate galaxies from nongalaxies. This method was chosen because of the difficulty in obtaining both types of data, the images and photometry, and the need for double confirmation of the classification to estimate the quality of the results with each sample. To deal with the data in the ZoA, class balancing methods were applied to account for the 1:13 number imbalance in the galaxy and nongalaxy data. 

Previous to \cite{Daza+2023}, only a few works had used ML techniques over the ZoA regions. \cite{Vavilova+revista} generated galaxy distributions and properties to compare the artificial survey with the real data in the region. \cite{Jones+2019arXiv} applied convolutional neural network (CNN) and evolutionary algorithms with VISTA and UKIDSS data to study the behavior of these tools in this region.  The results in this latter case are promising, with a good percentage of accuracy but the authors nevertheless suggest that results should always be visually checked for this particular area.

In addition to NIR, X-rays offer an excellent window onto extended sources in the ZoA thanks to the transparency of the MW for hard X-ray emission. 
\cite{Zang+2021} (hereafter, ZZW21) performed an automated classification of sources using ML techniques in the entire 4XMM-DR9 survey, which covers a large part of the sky, including the thin
disk of the Galaxy. These authors also used data from AllWISE \citep{WISE2010}, SDSS \citep{SDSS2000}, and LAMOST \citep{LAMOST2012} surveys at different wavelengths. Despite the increasing number and quality of data, the identification of extragalactic sources at lower galactic latitudes is challenging due to high levels of interstellar extinction and contaminating light from stars and galactic objects. Furthermore, it is worth noting that the surveys used in ZZW21 are all-sky surveys and do not concentrate on mapping at low latitudes, in contrast to the VVV survey. Our main goal is to compare the classifications of ZZW21 with the NIR data from the VVV survey, in order to provide quantitative validation of the ZZW21 classifications in these regions.

The paper is organized as follows: in section \S~\ref{TheData} we describe the data, tools, and procedures used in this work  to select the sources to be compared with the results of ZZW21. In section \S~\ref{inspeccion}, special attention is given to the visual inspection of the sources and in section \S~\ref{Analisis} we  provide specific examples and define {``interesting zones''} that can be used to understand the nature of the objects. 
Finally, in section \S~\ref{conclusion}, we present a discussion about the application of ML algorithms in the ZoA and outline our main conclusions.

%__________________________________________________________________
\section{The data} \label{TheData} 

The VVV survey provides NIR data in the southern Galactic disk. In recent years, we have used these data to identify and study galaxies behind the disk.  We ran \sex+\psfex\ \citep{PSFEx} on all the images, finding 177,838,607 sources, mainly Galactic objects, such as stars, star associations, groups, and forming star regions. The extragalactic sources have colors that are different from those of  Galactic objects and we used a methodology that selects galaxy candidates based on color cuts \citep{Baravalle+2018, Baravalle+2019}.  All the candidates were visually inspected using processed images available at the ESO Science Archive and the VISTA Science Archive\footnote{\url{http://horus.roe.ac.uk/vsa/}} \citep[VSA;][]{VSACross+2012}. The VVV near-IR Galaxy Catalog \citep[VVV NIRGC;][]{Baravalle+2021} is the result of this procedure, and is the largest catalog of galaxies in the southern Galactic disk. It consists of 5,563 visually confirmed galaxies behind the MW.

The European Space Agency’s X-ray Multi-Mirror Mission
\citep[XMM–\textit{Newton};][]{XMM+2001} was launched in 1999, making observations in the X-ray, ultraviolet, and optical passbands. The 4XMM survey \citep{4XMM+2020} has further improved our understanding of the X-ray Universe, providing a deep and detailed look at the X-ray sources in the sky. With extensive X-ray source catalogs, 4XMM has allowed astronomers to study a wide range of phenomena in the  $0.2-12\,\si{\kilo\eV}$ energy range, including X-ray binaries, supernova remnants, and X-ray-emitting stars. The typical positional accuracy is about 2 arcseconds.

\cite{Zang+2021}  employed ML techniques to classify galaxies in X-rays and included information from optical and IR in the training set.
The final sample from ZZW21 consists of 550,124 objects, each with a classification parameter indicating the probability that the object is a star, a galaxy or a quasi-stellar object (QSO). 
The probability for galaxy classification is determined by the ML algorithm employed in combination with the bands used. The settings with best accuracy for each sample are X-ray only (Rotation Forest, 77.80\%), X-ray/optical (LogitBoost, 92.82\%), X-ray/IR (Random Forest, 89.42\%), and X-ray/optical/IR (LogitBoost, 94.26\%). The minimum probability to be classified as a galaxy using only X-ray information with the Rotation Forest algorithm is $0.333$, which is one-third of the sample.
From the sample of ZZW21, there are 15,423 objects in the VVV disk region (between $295^{\circ} < \ell < 350^{\circ}$ and $-2^{\circ}<b<2^{\circ}$), which the authors classified  as stars, galaxies, and QSOs. As there are no galaxies in these regions obtained using optical data and only a few with IR in these regions, we decided to  use the classification and probabilities P$_X$ from only X-ray data.
There are 1,666 stars ($10.80\%$), 9,726 galaxies ($63.06\%$), and 4,031  QSOs ($20.14\%$), with an important imbalance, mainly between stars and galaxies. Hereafter, we refer to the subsample of 9,726 ZZW21 galaxies in the VVV regions of the galactic disk as the \textit{galXray} sample.
The median of their probabilities P$_X$ to be
galaxies is 0.664.  There are 4,829 sources with higher-than-median probabilities of being considered galaxies, which represent 49.65\% of the sample. Our main goal in this work is to detect these X-ray galaxies in the NIR passbands of the VVV survey.

\begin{figure}[H]
    \centering
    \includegraphics[width=0.5\textwidth]{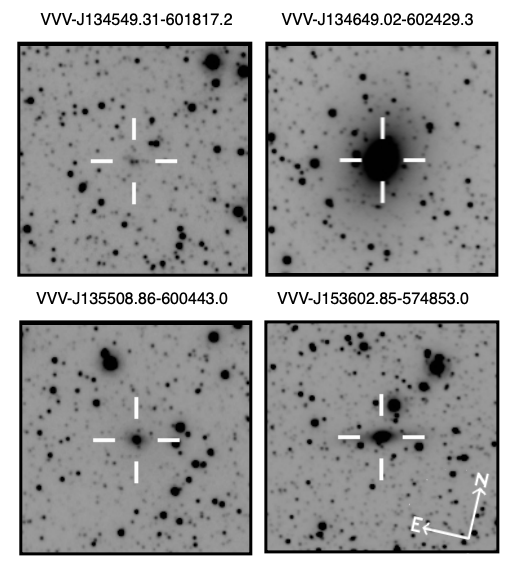}
    \caption{Galaxies in common between VVV NIRGC and the \textit{galXray} sample. The galaxies are shown in the K$_{s}$ VVV passband with 1$^{\prime}$ $\times$ 1$^{\prime}$ size. The orientation of all images is shown in the bottom-right panel.
    }
    \label{fig:true_gal}
\end{figure}

The \textit{galXray} sample was cross-matched with the VVV NIRGC catalog,  which revealed only four galaxies in common with differences in positions of smaller than 1.3 arcsec. The VVV NIRGC has only 45 galaxies in common with other authors (namely \cite{Schroder2007,Williams2014,Said2016,Schroder+2019a} (\citealt{Baravalle+2021}, Section 2.2)) and the \textit{galXray} sample has only one galaxy in common with \cite{Schroder2007}: DZOA 4653--11 (J134736.00-603703.8) with a CMB radial velocity of 4041 $\pm$ 86 km/s \citep{Radburn+2006} and a probability of P$_X = 0.552$ of being a galaxy from ZZW21.  Figure \ref{fig:true_gal} shows the VVV K$_{s}$ images  of the four galaxies in common between VVV NIRGC and the \textit{galXray}
sample, each of 1$^{\prime}$ $\times$ 1$^{\prime}$ in size. These are clearly early-type galaxies with probabilities P$_X$ of 0.575 and 0.561 (upper panels) and bulges with P$_X$ of 0.765 and 0.778 (lower panels). These objects have K$_s$ magnitudes brighter than 15.06 mag, as reported by \cite{Baravalle+2021}.

\begin{figure*}[t]
    \includegraphics[width=\textwidth]{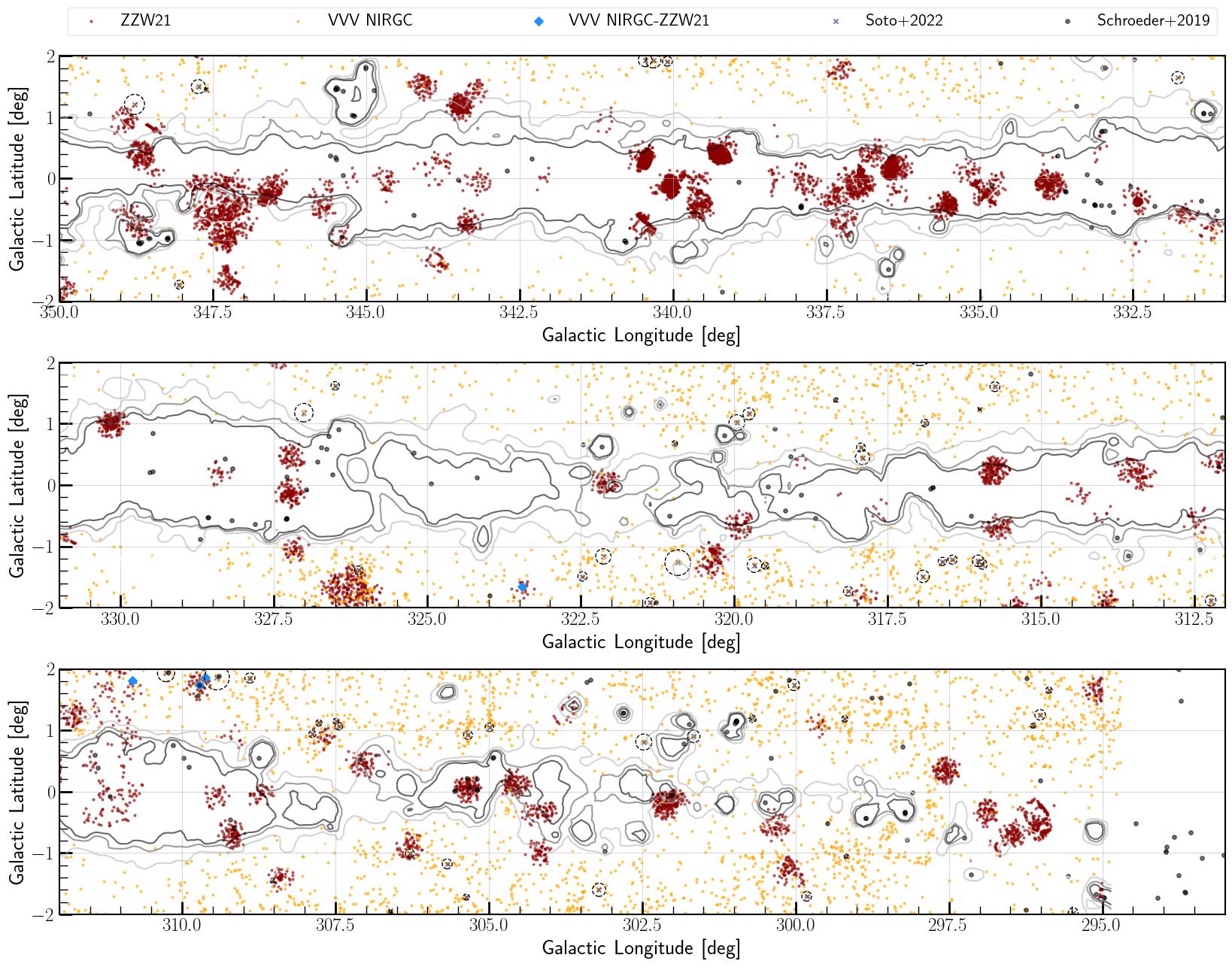}
    \caption{Distribution of galaxies from the ZZW21 in the VVV Southern disk region. The galaxies from \textit{galXray} sample are represented by red dots, the confirmed galaxies from the VVV NIRGC are in orange, the four galaxies in common between them are shown as  blue diamonds, and galaxies from \cite{Schroder+2019a} as black dots. The center of the overdensities reported by \cite{Soto+2022} are represented with a black ``x'' centered on a black dotted circle, which denotes the radius of each overdensity amplified by a factor of four. % and the positions of some important features of the LSS are also marked as %the \pame{typical} position of the Norma and Shapley superclusters and the Centaurus Wall, as well as the Great Atractor at the 2000 \citep{Kraan-Korteweg+2000}.
    %The black squares identify the studied \textit{Interesting Zones} (Section \ref{Analisis}).
    The $A_V$ isocontours derived from the extinction maps of \cite{Schlafly2011} are superimposed in a gray gradient with levels of 11, 15, 20, and 25 mag.}
    \label{fig:map_eross}
\end{figure*}

We also performed a cross-match, with a radius of 1.3 arcsec, between the \textit{galXray} sample with the original output of \sex+\psfex\ in the Galactic disk to see the morphology of the sources.  We  found 3,229 sources in common, of which 3,225 are Galactic sources. Of these, 3,183 are point sources and 64 extended ones, mainly consisting of gas clouds. Figure \ref{fig:map_eross} shows the distribution of galaxies within the VVV Southern disk area obtained from the \textit{galXray} sample, the VVV NIRGC, and \cite{Schroder2007}.
In general, the \textit{galXray} sample is distributed evenly throughout the inner regions of the disk and exhibits high extinctions at very low Galactic latitudes. On the other hand, the VVV NIRGC is located at higher latitudes. There is no overlap between the surveys due to the high  interstellar extinctions and the differing wavelengths of each survey.  
Also shown are the overdensities present in the area introduced by \cite{Soto+2022} and the $A_V$ isocontours derived from the extinction maps of \cite{Schlafly2011}.

\begin{figure}%[H]
    \centering
    \includegraphics[width=.4\textwidth]{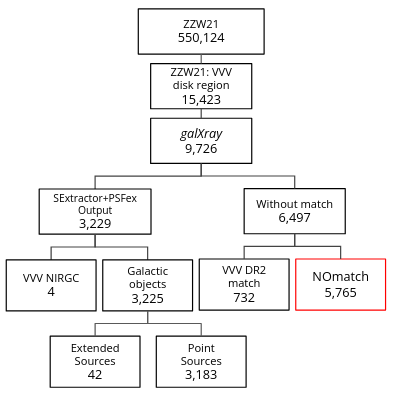}
    \caption{Flow-chart showing the  selection of sources from the ZZW21 sample making up the \textit{NOmatch} sample, which is our main concern in this work.}
    \label{fig:diagrama}
\end{figure}

The remaining 6,497 sources with no counterpart in the \sex+\psfex\, catalog, which constitute 66.8\%  of the \textit{galXray} sample, underwent cross-matching with all surveys available in VizieR.  This cross-match yielded 732 sources in common with the VVV-DR2 survey, which are predominantly bright objects, specifically stars. %Our methodology to find galaxies had previously removed them in order to recognise galaxies \citep{Baravalle+2018}.
Additionally, the remaining 5,765 sources did not yield any results.  Furthermore, these sources lack counterparts in all wavelengths, including the Gaia-DR3 survey \citep{GAIAmission+2016}. This sample is referred to as the \textit{NOmatch} sample throughout this paper.  Figure \ref{fig:diagrama} summarizes the procedure adopted to select the final sample to identify the different  sources  that are part of the \textit{galXray} sample of galaxies in the ZZW21 study.\\

%__________________________________________________________________

\section{Visual inspection} \label{inspeccion}

To improve our understanding of the sources in the \textit{NOmatch} sample, we visually  inspected the 5,765 sources using $30^{\prime \prime} \times 30^{\prime \prime}$ $Z$, $Y$, $H$, $J$, and $K_s$ stamps centered at the position of the sources in the VVV-DR5 images. %Initially, we conducted a quick visual assessment, identifying 15 field types for each source in each band. 
The inspection involved observing differences in surface brightness of the sources in the five VVV NIR passbands as we did for the first time in \cite{Baravalle+2018} for galaxy identification and classification.
Stellar objects exhibit comparable surface brightness in all five VVV passbands, whereas extended sources possess larger surface brightness at longer wavelengths (\textit{J, H,} and \textit{$K_s$}), but are faint or barely detectable at shorter wavelengths (\textit{Z} and \textit{Y}). This inspection considered both the central source and its surroundings in order to identify and characterize common features. These features are not necessarily mutually exclusive, which means a single source can have multiple associated features. To improve the overall study, we kept the primary classification. The purpose of this inspection was to validate the galaxy classification of ZZW21. The sources detected by XMM--\textit{Newton} might not be observable in the NIR of the VVV survey because of the severe conditions of the Galactic disk with high interstellar extinction and stellar contamination. Our inspection allowed us to divide the sample into ten distinct groups or categories.

Here, we provide a brief description of the various features found in the NIR images of the southern Galactic disk region in order of the number of cases found.
The first and most common are normal "crowded regions" and "stars".  At lower latitudes, the presence of the Galactic disk  causes the fields to be heavily contaminated by stars.  Together they account for 73.7\% of the features found in the \textit{NOmatch} sample.  As noted above, the stars have approximately  the same surface brightness regardless of the passband. No extended structures are observed in the frames. %\textbf{STAR}
%This case, which corresponds to 13.4\%, consists of a star, whether it is small, faint, or a large, centrally located star in the frame. This category is not mutually exclusive, as cases with saturated stars, spikes, or star clusters can also be observed.
"Empty central regions" refer to areas in the center that lack detectable sources in the NIR passbands. %With a total of 13.4\% of sources classified in this way, it corresponds to an "apparently empty" zone in the center, without discernible stars near the central area or any changes indicating the presence of galaxies through dust as we move into the NIR. It is essential to note that the term "Empty-Central Region" does not imply complete emptiness but rather indicates the absence of detectable objects at the telescope's VVV point.
 %In the case of a ``\textit{Saturated Star}'', a very bright star is present somewhere in the frame, which is saturated in one or several bands. This star is characterized by its larger size compared to other sources and displays a white point at its center. Generally, this classification is accompanied by the \textit{spikes} observed in the previous category.%%%%%Cuando hice la inspección visual, deje igual las estrellas saturadas en el borde, varias de ellas presentan spikes tambien y ahi fue cuando le preguntamos a nicola si en esas igual podia afectar la deteccion en X-ray
"Star associations", "Saturated bright stars", and "spikes" represent less than 5\%  of the cases.  The star associations feature comprises up to four stars located in close proximity to each other in or near the stamps' central regions. Bright stars might be present either at the center of the stamps or in other areas, and could significantly contaminate the  objects.  The saturated stars with larger radii and spikes from nearby stars could strongly affect the surrounding area.   %In certain bands, the clustering of stars can be easily confused with a strong extragalactic source due to their proximity and levels of stellar extinction. This case accounts for 6.2\% of the cases and can often be found alongside other classifications, such as "Empty-Central Region" or "Saturated Bright Star." \
The spikes display a diffraction pattern of a massive nearby star on the stamps.  % in the form of \textit{spikes}. For this category, the presence of this diffraction pattern alone was considered for classification. 
Less than 1\% of the cases correspond to star-forming regions, which are typically found in the disks of spiral galaxies.  We include the "star-forming regions" observed
in most passbands with no variations in the star brightness. %Its change is not as significant as in the case of PUVSFR, and it is possible to notice other types of structures, such as groups of stars, saturated stars, or weak "Spikes" from stars outside the frame. This case accounfts for 1.1\% of the total sample.
We also define "photometrically ultravariable stars" (PUVS), which refers to the presence of a star near the center that varies consistently in brightness between passbands. Although sometimes corresponding to variable stars, these objects are occasionally misidentified as extended sources.  %Here, the group of stars can easily be mistaken for extended sources when galactic extinction levels are high, as in this region.
We also designate  "photometrically ultravariable star-forming regions" ({PUVSFR}), which show a significant high surface brightness levels at longer wavelengths (J, H, and $K_s$), often spanning a large portion of the frame and even with the presence of saturated stars. At shorter wavelengths, these appear to be made up of gas clouds surrounding the stars.% or the frame is not easily visible or, in some cases, cannot be detected. 
\, In 13 cases, we also found bright star associations characterized by several massive stars, mostly saturated and displaying spikes. These massive stars are close in the sky suggesting that they belong to the same highly star-forming region with high X-ray emission.\\

\begin{figure}
    \centering
    \includegraphics[width=0.5\textwidth]{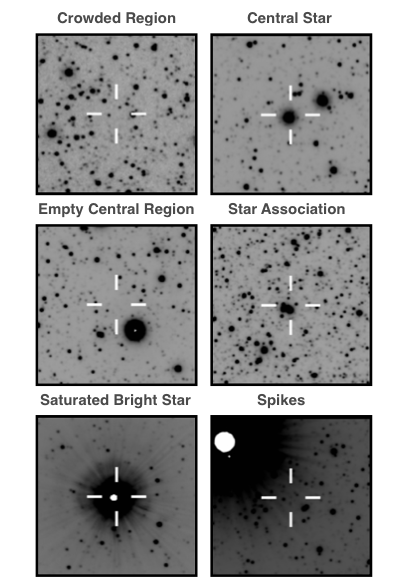}
    \includegraphics[width=0.5\textwidth]{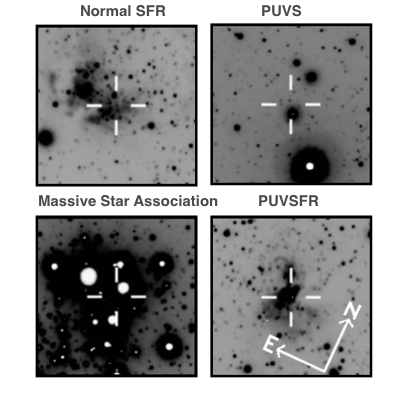}
    \caption{Examples of objects found in the \textit{NOmatch} sample are shown in the VVV K$_{s}$ passband with the size of 1$^{\prime}$ $\times$ 1$^{\prime}$. The orientation of all images is shown in the bottom-right panel.}
    \label{fig:examples_class}
\end{figure}

Table \ref{tab:nan} summarizes the features found in the stamps of the \textit{NOmatch} sample together with the number of occurrences and the percentage relative to the total number of sources. Figure \ref{fig:examples_class} shows some examples in the VVV K$_{s}$ passband of the different features. The ZZW21  galaxies with probabilities P$_X$ higher than 0.95 in the \textit{galXray} sample were visually inspected and all the objects are bright stars.  This represents the 9\% of the sample. Upon request, the samples of objects with different features can be provided.  

\begin{table}%[H]
    \centering
    \begin{tabular}{llrr}
    \hline \hline
    Features in the NIR & Total number & \% \\
    \hline
    Crowded region           & 3,243 & 56.25\%\\
    Central star             & 1,008 & 17.48\%\\
    Empty central region     & 971   & 16.84\%\\
    Star association         & 266   & 4.65\%\\
    Saturated bright star    & 153   & 2.61\%\\
    Spikes                   & 48    & 0.83\%\\
    Normal SFR               & 29    & 0.50\%\\
    PUVS                     & 22    & 0.38\%\\
    Massive star association & 13    & 0.23\%\\
    PUVSFR                   & 12    & 0.21\%\\
%    \hline
%    Special Case            & 2     & 0.03\%\\
    \hline \hline
    \end{tabular}
    \caption{Features found in the NIR images through visual inspection.}
    \label{tab:nan}
\end{table}

\section{Detailed inspection of interesting zones} \label{Analisis}

\begin{figure*}[t]
    \centering
    \includegraphics[width=\textwidth]{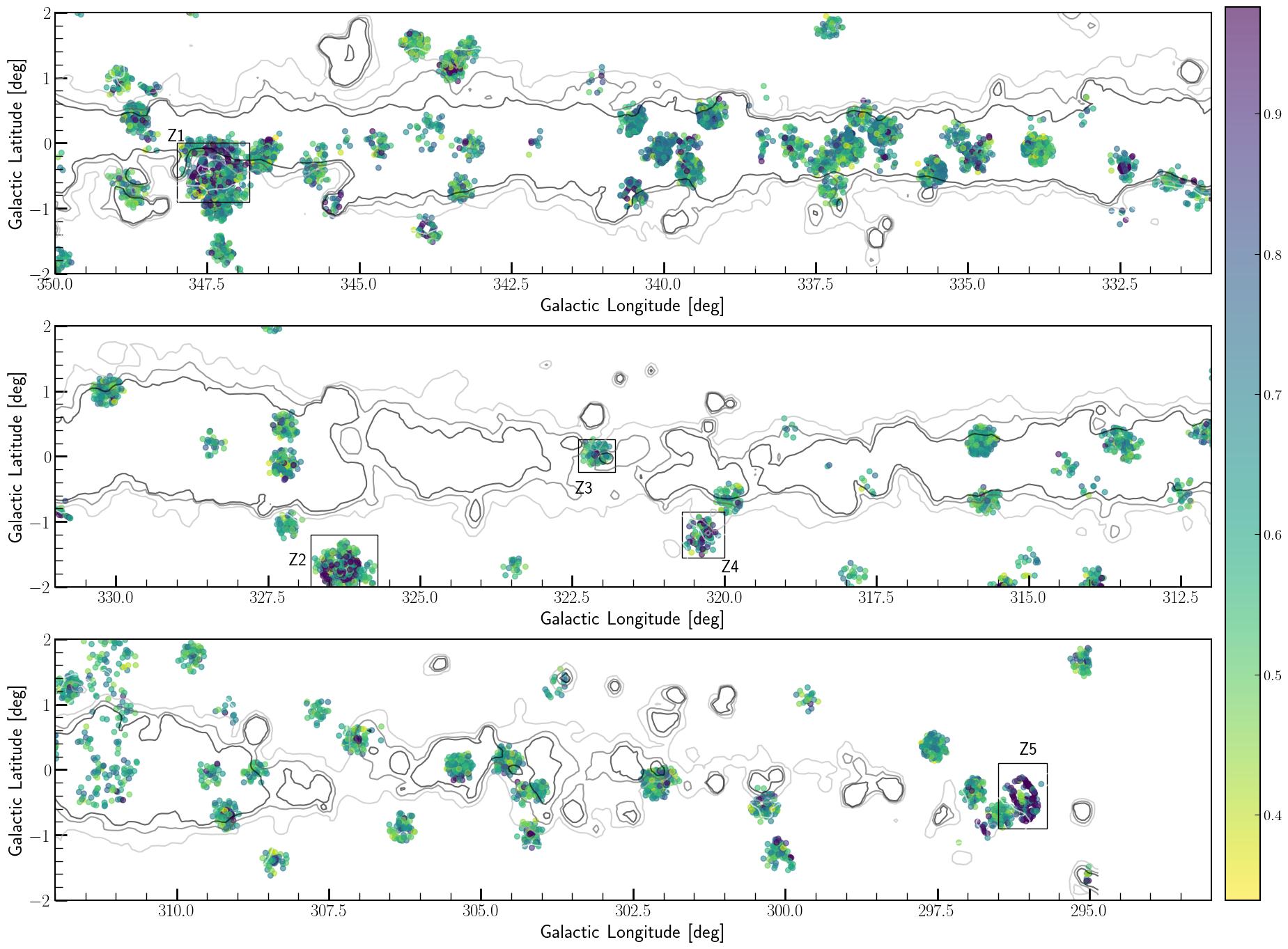}
    \caption{Distribution of sources classified as galaxies in the southern Galactic disk of the VVV survey. The galaxies from \cite{Zang+2021} are color-coded according to their probability P$_X$ of being a galaxy. The black squares shows the "interesting zones" studied. The $A_V$ isocontours derived from the extinction maps of \cite{Schlafly2011} are superimposed in gray scale with levels of 11, 15, 20, and 25 mag.}
    \label{fig:map_Zprob}
\end{figure*}

To study the distribution of the X-ray sources in the \textit{galXray} sample at lower Galactic latitudes, we defined some {interesting zones} inspired by the patterns visible in Figure \ref{fig:map_eross}. We identified the 9,726 sources classified as galaxies from the \textit{galXray} sample, the VVV NIRGC, and \cite{Schroder+2019a}.  
We now focus our interest on the distribution of the sources from the \textit{galXray} sample represented in red. We selected five {interesting zones}, 
Z1, Z2, Z3, Z4, and Z5. These are distributed across the Galactic plane, each one with distinct characteristics and showing a high concentration of X-ray sources. Figure \ref{fig:map_Zprob} shows the distribution of sources classified as galaxies color-coded according to the probability P$_X$ of being a galaxy, as defined by ZZW21. Most of the regions exhibit high interstellar extinction, particularly Z3, whilst Z5 has the lowest extinction in comparison.  Furthermore, Z1 to Z4 are located in the Norma Supercluster region \citep{WoudtKraan2001}, which is of major importance for the LSS and the Great Attractor.

\begin{figure*}
\centering
\includegraphics[width=0.30\textwidth,height=0.28\textwidth]{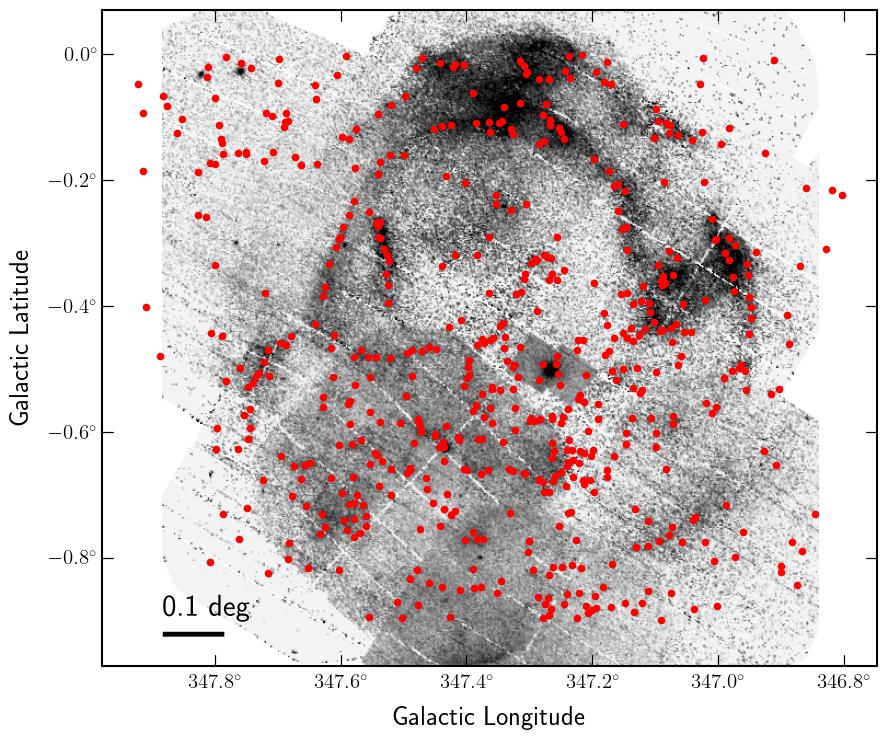}
\includegraphics[width=0.30\textwidth,height=0.28\textwidth]{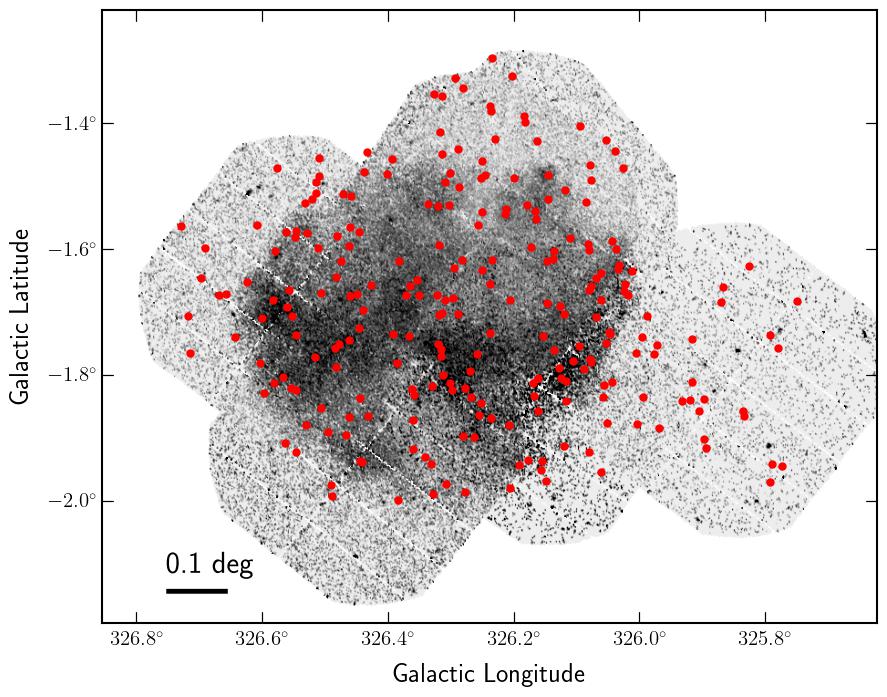}
\includegraphics[width=0.30\textwidth,height=0.28\textwidth]{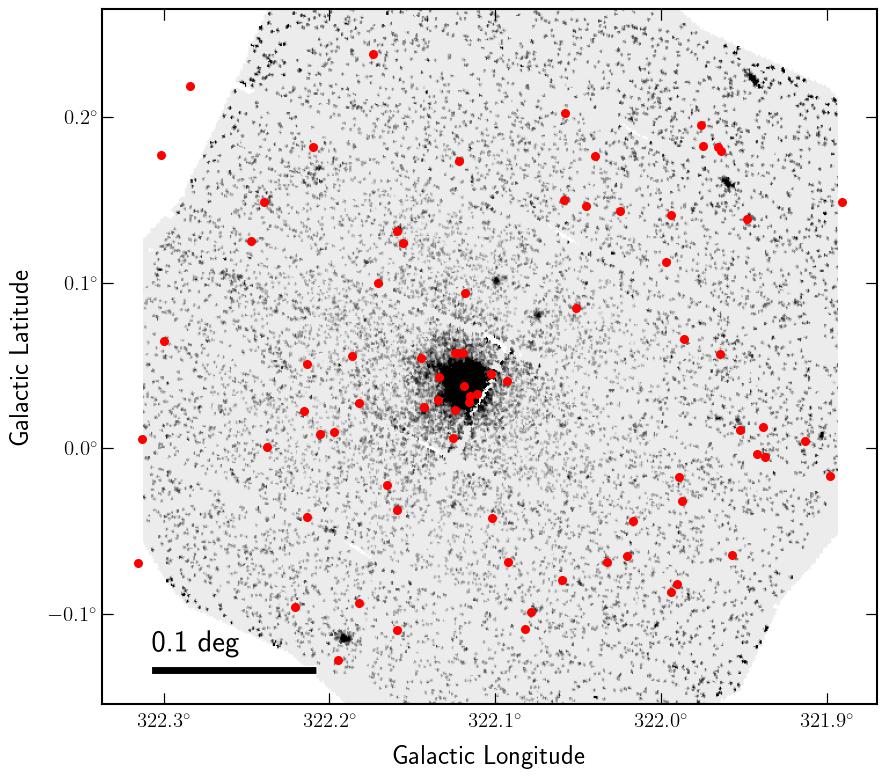}\\
\includegraphics[width=0.30\textwidth,height=0.28\textwidth]{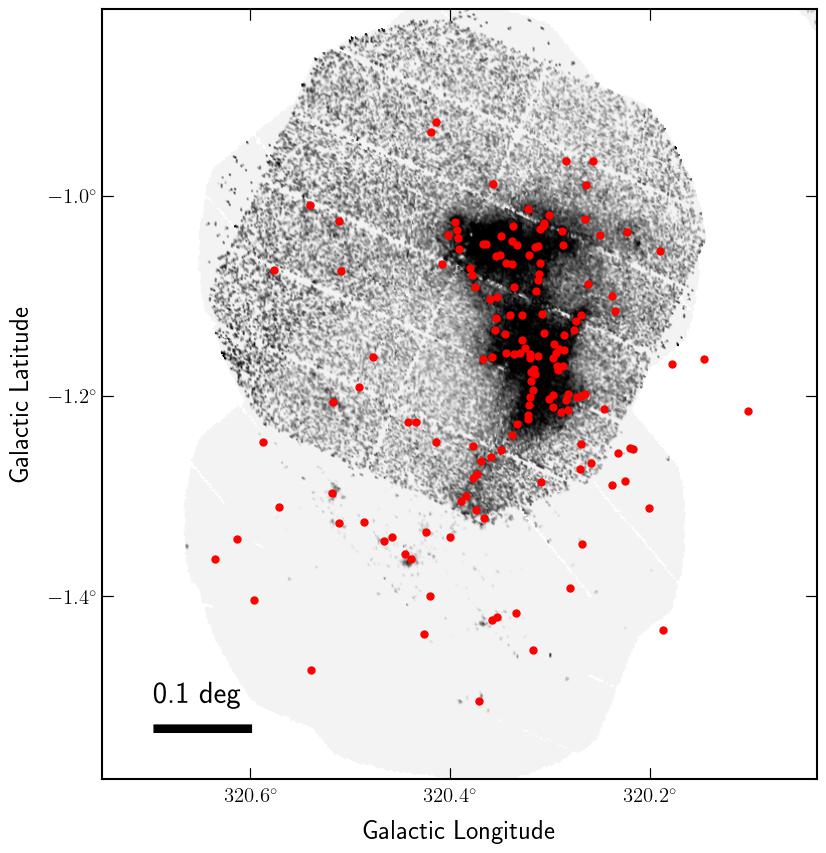}
\includegraphics[width=0.30\textwidth,height=0.28\textwidth]{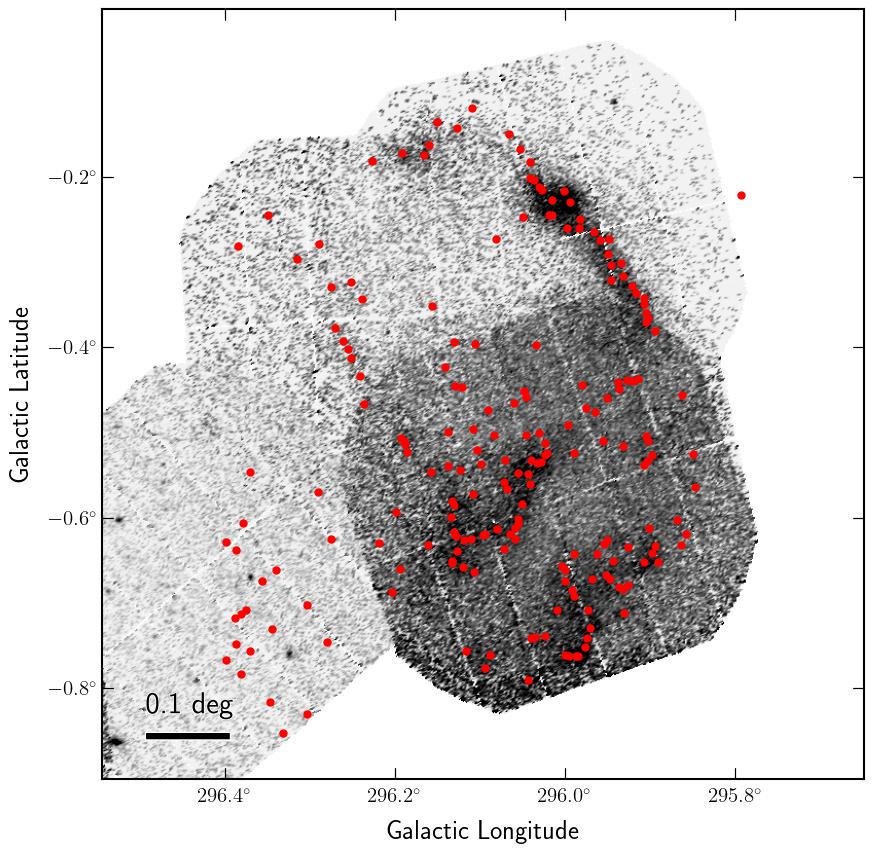}
\caption{From top-left to bottom-right, the XMM eb3 images of the Z1 to Z5 {interesting zones}, respectively, including the \textit{galXray} objects of ZZW21 in each region in red dots.}
\label{fig:Zonas_withdata}
\end{figure*}

Figure \ref{fig:Zonas_withdata} shows the X-ray images in the eb3 channel ($1.0-2.0$ $\si{\kilo\eV}$) from the XMM--\textit{Newton} Observatory of the interesting zones Z1 to Z5  with the \textit{galXray} sources highlighted as red points in each specific region. This passband helps us to verify the existence of extended Galactic structures because it is more affected by the presence of strong interstellar absorption and is less contaminated by hard X-ray emission. A peculiar Galactic extended structure resembling a bubble can be identified in some of these images. In the  Z1, Z2, and Z5 zones, the symmetrical shape suggests a supernova remnant (SNR) with the additional presence in X-rays of a central point source, as in the case of Z1. Conversely, Z4 seems more indicative of a star-forming region, with the gas distribution concentrated in one specific area.  It is also observed that the sources of the  \textit{galXray} sample present a nonuniform distribution, favoring the structure of the hot gas in each zone.  An exception is Z3,  where the distribution of the sources shows a nearly homogeneous distribution throughout the area.

Zones Z1 to Z5 were also observed at different wavelengths, including radio, mid-infrared (MIR), NIR, and optical, using images provided by the Sydney University Molonglo Sky Survey \citep[SUMSS;][]{SUMSS+1999}, AllWISE \citep{WISE2010, NEOWISE+2011}, VVV, and the Supercosmos H-alpha Survey \citep[SHS;][]{SHS+2005}, respectively. 
Figures \ref{fig:Z1_aladin} to \ref{fig:Z5_aladin} show the {interesting zones} at the most relevant wavelength of each survey.

%%%%%%%%%%%%%%%%%%%%%%%%%%%%%%%%%%%%%%%%%%%%%%%%%%%%%%%%%%

\subsection{Supernova remnants}
In the Z1 region, centered at $\alpha$: 17h13m21.31, $\delta$:-39d41m32.05, it is possible to observe a symmetrical circular structure in the XMM images that resembles a sphere surrounding a central point source. We note that 562 \textit{galXray} objects in this region (red points; see upper left panel of Fig. \ref{fig:Zonas_withdata}) follow the X-ray emission structure, concentrating over the darker regions.  According to different studies \citep{Slane+1999,Vasquez+2005, Cassam-Chena+2004,Tateishi+2021}, this region corresponds to the shell-type SNR RX J1713.7–3946, which is characterized by nonthermal emission. Its central X-ray point source is 1WGA J1713.4–3949, which has an X-ray--to--optical flux ratio that is consistent with that of a neutron star. Another possibility is that the source is an extragalactic background source  \citep{Slane+1999}, and therefore it is highly probable that the central point source seen in X-rays is the compact relic of the supernova progenitor of the remnant in the category of type II supernovae \citep{Cassam-Chena+2004}.
The other point source 12 arcmin away from the central X-ray source corresponds to the red super giant star HD 155603.

In the Z3 region, centered at $\alpha$: 15h20m34.605s, $\delta$: -57d07m56.599s), 
the central source shown in the upper right panel of figure \ref{fig:Zonas_withdata} corresponds to the neutron star X-ray binary Circinus X-1 within a SNR, known as one of the brightest X-ray sources in the sky. \cite{Heinz+2013} studied the natal SNR of the accreting neutron star Circinus X-1, comparing the emission in X-ray with radio. In the SUMSS stamp (radio) in Figure \ref{fig:Z3_aladin}, the SNR and the jet of the binary source \citep{Sadeh+1979,Phillips+2007,Johnston+2016,Coriat+2019}  can be seen at the center.\\

The \textit{galXray} sample in the Z1 and Z3 regions provides a clear delineation of the gas structure in the SNRs. The sources in these two regions have median probabilities of being galaxies  P$_X$  of 0.712 and 0.674, respectively.

\subsection{Other regions}

In the Z2 ($\alpha$: 15h52m26.793s, $\delta$: -56d11m29.343s), we observe that the \textit{galXray} emission data (red points, figure \ref{fig:Zonas_withdata}) follow the distribution of the hot gas circular structure for XMM and SUMSS images. For Z4 ($\alpha$: 15h14m43.7601s, $\delta$: -59d09m40.265s), nonuniform structure is found, both in the eb3 channel of XMM and other passbands (Figures \ref{fig:Zonas_withdata} and \ref{fig:Z4_aladin}). In Z5 ($\alpha$: 11h51m50.437s, $\delta$: -62d35m19.304s), and the red points of the \textit{galXray} sample in Figure \ref{fig:Z5_aladin} follow part of a double-shell structure, which is also observed in the SUMSS passbands.
At present, no additional data are available in the literature for these areas.

%______________________________________________________________
\begin{figure*}
    \centering
    \includegraphics[width=\textwidth]{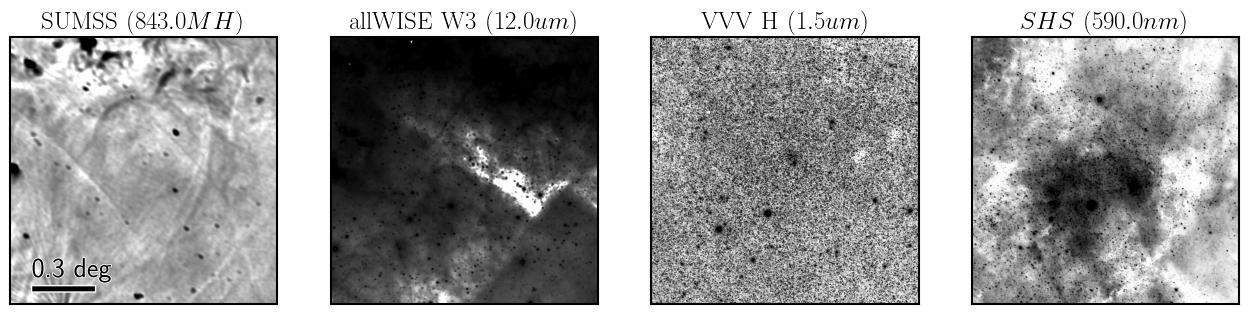}
    \caption{Stamps at different wavelengths for the Z1 interesting zone with available surveys.}
    \label{fig:Z1_aladin}
\end{figure*}
\begin{figure*}
    \centering
    \includegraphics[width=\textwidth]{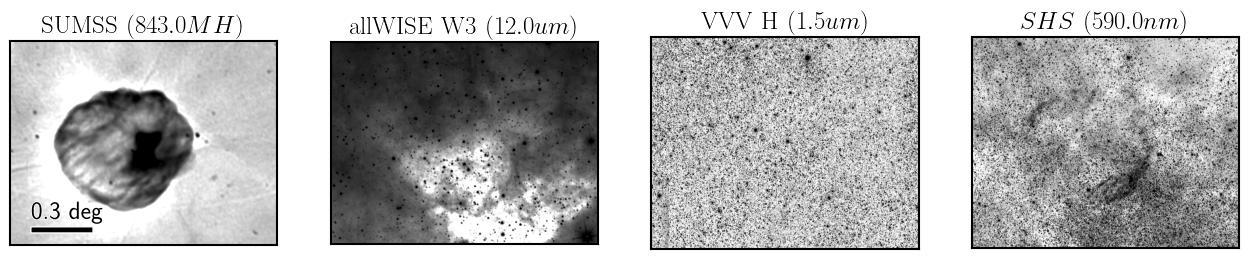}
        \caption{Same as Fig. \ref{fig:Z1_aladin}, but for Z2 region.}
    \label{fig:Z2_aladin}
\end{figure*}
\begin{figure*}
    \centering
    \includegraphics[width=\textwidth]{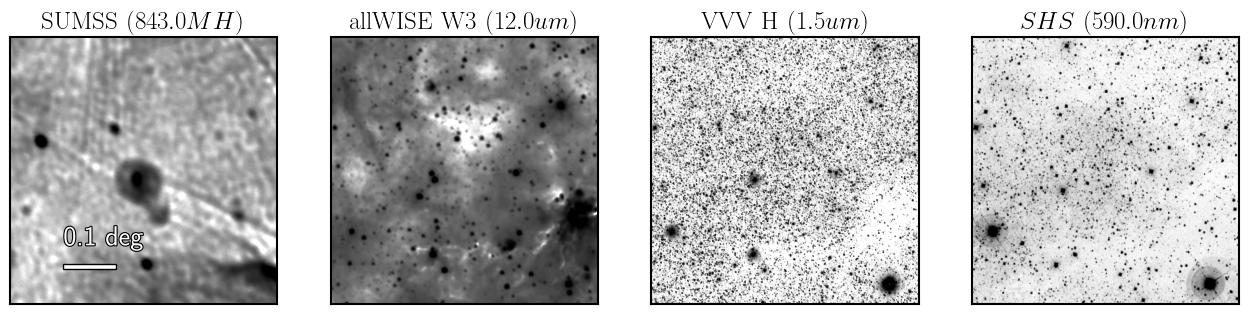}
        \caption{Same as Fig. \ref{fig:Z1_aladin}, but for Z3 region.}
    \label{fig:Z3_aladin}
\end{figure*}
\begin{figure*}
    \centering
    \includegraphics[width=\textwidth]{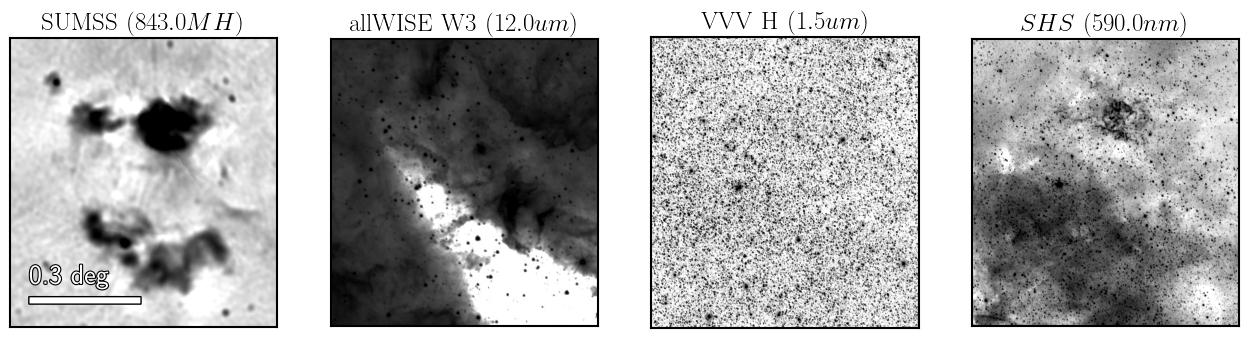}
            \caption{Same as Fig. \ref{fig:Z1_aladin}, but for Z4 region.}
    \label{fig:Z4_aladin}
\end{figure*}
\begin{figure*}
    \centering
    \includegraphics[width=\textwidth]{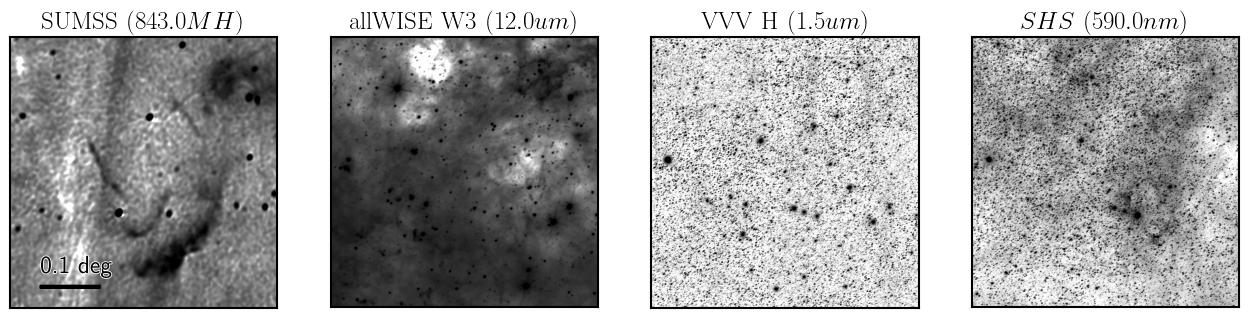}
            \caption{Same as Fig. \ref{fig:Z1_aladin}, but for the Z5 region.}
    \label{fig:Z5_aladin}
\end{figure*}

Table \ref{tab:median} summarizes the statistics of the probabilities that the sources are galaxies from the work of ZZW21 for the {interesting zones}, including those from  all-sky XMM data and \textit{galXray} sample for comparison.  The table shows the identification and the number of objects in each sample in columns (1) and (2) and the quartiles in the columns (3) to (5). The sources in Z2, Z4, and Z5 have probabilities P$_X$ of being galaxies with median values of 0.649, 0.744, and 0.966, respectively.  The median of  Z5 is the highest and is a very different value compared with the other regions.
Also, it is somewhat puzzling that the quartile distribution of the classification probability in the ZZW21 and \textit{galXray} samples  does not show significant changes between the all-sky and the Galactic disk results, despite the different extinction levels.

\section{Discussion and final remarks} \label{conclusion}

Classifying extragalactic objects, especially through automated methods, is very challenging at lower Galactic latitudes due to high interstellar extinction and the small number of galaxies compared with stars. \cite{Zang+2021} used machine learning algorithms and defined the training set using X-ray data. These authors presented an all-sky catalog of galaxies with associated probabilities. The ML algorithms might be deficient for the detection and classification of galaxies that are outside the range of characteristics used in the training models. In this work, we compared the galaxies found by these latter authors at energetic wavelengths with the objects obtained by \sex+\psfex\, in the NIR regime.

\begin{table}[H]
    \centering
    \begin{tabular}{lcccc}
    \hline \hline
    Galaxy  & \#      & Q1    & Q2    & Q3    \\ 
    Sample  & objects &       &       &       \\ 
    \hline
    ZZW21   & 191,528 & 0.538 & 0.657 & 0.804 \\
    galXray & 9,726   & 0.552 & 0.664 & 0.743 \\
    Z1      & 562     & 0.591 & 0.712 & 0.953 \\
    Z2      & 244     & 0.539 & 0.649 & 0.958 \\
    Z3      & 78      & 0.552 & 0.674 & 0.712 \\
    Z4      & 158     & 0.566 & 0.744 & 0.963 \\
    Z5      & 203     & 0.704 & 0.966 & 0.972 \\
    
    \hline  \hline
    \end{tabular}
    \caption{Statistics of the probabilities P$_X$ for different samples and zones.}
    \label{tab:median}
\end{table}

\cite{Zang+2021} classified galaxies using the Rotation Forest method, relying solely on X-ray data. However, data from other wavelength ranges were incorporated during the training phase, which introduced emissions with potentially distinct signals. It is important to double check with a different data set, such as the VVV survey in the NIR, because the use of diverse data sources may lead to misclassifications, especially at lower Galactic latitudes. In such cases, it is crucial to consider data transfer or transfer learning when performing classifications. This strategic approach allows knowledge gained from one data set to enhance the algorithm's performance on another, potentially mitigating challenges associated with varied signal and noise characteristics. In the case of ZZW21, the algorithm was trained using SDSS data that mainly consist of bright and large galaxies. No optical galaxies were found in the studied region. It is possible that the algorithm may classify objects with X-ray emission as galaxies, but this may only be valid when they are bright and not obscured by Galactic dust.
Hence, even objects with X-ray emission resembling that of SDSS galaxies should be treated with caution. In the samples through the Galactic disk, certain objects may be erroneously classified as galaxies, when in reality they are blended stars or other extended Galactic structures.

From the all-sky 4XMM-DR9 survey, there are 15,423 objects from ZZW21 within the VVV southern Galactic disk area. In this sample, ZZW21 classified 1,666 stars, 4,031 quasars, and 9,726 galaxies. In this region, \cite{Baravalle+2021} obtained the VVV NIRGC catalog with 5,563 visually confirmed galaxies. The cross-match between these two samples results in only four galaxies in common.  There is also one galaxy reported by \cite{Schroder2007} in this region. 

The 5,765 objects of the \textit{NOmatch} sample were visually inspected in this work. This was the most critical part, which included regions of high interstellar extinction and stellar crowding. It is evident that the majority of the ten different features summarized in Table \ref{tab:nan} correspond to Galactic regions normally crowded with several stars distributed over the whole area. Around 25\% of the cases exhibit a bright or saturated star at the center of the stamp. The cases of \textit{Normal Crowded Region} and \textit{Empty Central Region} account for 73\% of all cases. The absence of galaxies in the NIR regime does not necessarily imply a lack of galaxies in these regions, but rather suggests that it may be challenging to detect them, especially in regions with higher interstellar extinctions. The galaxies can be observed with the XMM but remain undetectable in the NIR passbands. Hence, the most significant result is the lack of galaxies in these regions.\\

We also used the results of several surveys at different wavelengths to perform a visual panchromatic inspection. We defined five ``{interesting zones''} based on the data distribution from ZZW21. Images of each area were compared at different wavelengths, from radio to X-rays (Figures \ref{fig:Z1_aladin} to \ref{fig:Z5_aladin}). They also show high probabilities of being a galaxy, as seen in Figure \ref{fig:map_Zprob} and Table \ref{tab:median}. Based on these images and previous results, we might conclude that these {interesting zones} correspond to the emission from  extended Galactic events rather than individual galaxies.
The sources classified as galaxies by ZZW21 belonging to the {interesting zones} are part of extended Galactic structures, such as SNRs or star-forming regions. Further studies in these regions are needed.  

Imbalanced datasets are frequently encountered in ML and pattern recognition \citep{Lemaitre}, compromising the learning process.  Most of the standard ML algorithms expect balanced class distribution or an equal misclassification cost \citep{He_Garcia_2009}. This problem can be critical when trying to distinguish  galaxies from nongalaxies  at lower Galactic latitudes where the numbers of stars and associations are extremely important. Conventional classifiers often prioritize the minimization of the overall error rate, which can lead to a bias towards the majority class, resulting in an inaccurate classification. The most common metric used in classification is accuracy, which measures the ratio of correct predictions to the total number of input samples \citep{Blagec+2020}. In the zone of avoidance, when dealing with imbalanced data sets, which are often found when classifying galaxies and nongalaxies, accuracy alone may not be the optimal metric. In \cite{Daza+2023}, we used the F1-score as a better option, as it considers both the quantity and quality of the classification.  Our classification algorithms were centered on the 5,509 VVV NIRGC galaxies and 74,238 nongalaxies in the southern part of the Galactic disk using the VVV survey and the VVV NIRGC. We  considered regions with varying Galactic extinction levels, employing both the CNN method with NIR images and the XGBoost method with photometric and morphological VVV NIRGC data. These two samples were used as training sets to separate galaxies from nongalaxies in the northern Galactic disk using the VVVX survey, taking into account the number imbalance present in the dataset.

Our work highlights the importance of having a representative training set when working on the ZoA using ML. An appropriate training set ensures accurate and reliable classification, improving the purity of the positive class in particular, and the results in general, which is of paramount importance when mapping the LSS at lower galactic latitudes.  

\begin{acknowledgements}
      We would like to thank the anonymous referee for useful comments and suggestions which have helped to improve this paper.
      P. M. C. thanks the support of the Universidad de La Serena and the Southern Office of Aerospace Research and Development of the Air Force Office of the Scientific Research International Office of the United States (SOARD/AFOSR). J. L. N. C. is grateful for the financial support received from SOARD/AFOSR through grants FA9550-18-1-0018 and  FA9550-22-1-0037. M. V. A., L. B. and C. V. thank the support of the Consejo de Investigaciones Cient\'ificas y T\'ecnicas (CONICET) and Secretar\'ia de Ciencia y T\'ecnica de la Universidad Nacional de C\'ordoba (SeCyT). F. M. C. thanks the support of ANID BECAS/DOCTORADO NACIONAL 21110001. D.M. gratefully acknowledges support by the ANID BASAL projects ACE210002 and FB210003 and by Fondecyt Project No. 1220724. The authors gratefully acknowledge data from the ESO Public Survey program IDs 179.B-2002 and 198.B-2004 taken with the VISTA telescope, and products from the Cambridge Astronomical Survey Unit (CASU). 
      
\end{acknowledgements}

%-------------------------------------------------------------------
\bibliographystyle{aa} % style aa.bst
\bibliography{refs} % your references Yourfile.bib

\end{document}